\documentclass[twocolumn,showpacs,prl]{revtex4}

\usepackage{booktabs}
\usepackage{amsmath,amsfonts,amssymb}
\usepackage{txfonts}
\usepackage{type1cm}
\usepackage{graphicx}

\begin{document}

\bibliographystyle{apsrev}

\title{Experimental Decoy State Quantum Key Distribution with Unconditional Security Incorporating Finite Statistics}

\author{Jun Hasegawa$^{1,2}$, Masahito Hayashi$^{1}$, Tohya Hiroshima$^{1,3}$, Akihiro Tanaka$^{4}$, and Akihisa Tomita$^{1,3}$}
\affiliation{
$^{1}$ Quantum Computation and Information Project, ERATO-SORST, Japan Science and Technology Agency,\\
Daini Hongo White Building 201, 5-28-3 Hongo, Bunkyo-ku, Tokyo 113-0033, Japan.\\
$^{2}$ Department of Computer Science,
Graduate School of Information Science and Technology, 
the University of Tokyo,\\
7-3-1 Hongo, Bunkyo-ku, Tokyo 113-0033, Japan.\\
$^{3}$ Nanoelectronics Research Laboratories, NEC Corporation,\\
34 Miyukigaoka, Tsukuba 305-8501, Japan. \\
$^{4}$ System Platforms Research Laboratories, NEC Corporation,\\
1753 Shimonumabe, Nakahara-ku, Kawasaki 211-8666, Japan.
}

\date{}
%%%%%%%%%%%%%%%%%%%%%%%%%%%%%%%%%%%%%%%%%%%%%%%%%%%%%%%%%%%%%%%%%%%%%%%%%%%%%%%
\begin{abstract}
We propose the improved decoy state quantum key distribution incorporating finite statistics due to the finite code length and report on its demonstration.
In our experiment, four different intensities including the vacuum state
for optimal pulses are used and the key generation rate of 200 bps is
achieved in the 20 km telecom optical fiber transmission keeping the
eavesdropper's mutual information with the final key less than $2^{-9}$.
\end{abstract}
\pacs{03.67.Dd, 03.67.Hk, 03.67.-a}
%%%%%%%%%%%%%%%%%%%%%%%%%%%%%%%%%%%%%%%%%%%%%%%%%%%%%%%%%%%%%%%%%%%%%%%%%%%%%%%
\maketitle
%%%%%%%%%%%%%%%%%%%%%%%%%%%%%%%%%%%%%%%%%%%%%%%%%%%%%%%%%%%%%%%%%%%%%%%%%%%%%%%

Quantum key distribution (QKD) was originally proposed by Bennett and Brassard in 1984 \cite{BB84} as a protocol, by which two parties, Alice and Bob, share secret keys by using a quantum communication channel as well as a public classical channel \cite{QKD_Review}.
A remarkable feature is its unconditional security \cite{Security}; it is guaranteed by the fundamental laws of quantum mechanics and thereby QKD provides the unconditionally secure communication system.
In the practical setting of optical communication, however, it is the almost only option to substitute qubits in the original BB84 QKD protocol with heavily attenuated laser pulses because the perfect single photon emitting devices are not available in the current technology.
Such laser pulses - the phase randomized weak coherent states - contain inevitably the multiphoton states at small but finite probability, which give a malicious eavesdropper (Eve) a chance to obtain some amount of information on the shared keys by a photon-number-splitting attack \cite{PNS}.
Gottesman-Lo-L\"{u}tkenhaus-Preskill (GLLP) showed, however, that it is still possible to obtain unconditionally secret key by BB84 protocol with such imperfect light sources, although the key generation rate and distances are very limited \cite{GLLP}.

The recently proposed decoy state method \cite{Hwang,Wang,LMC,Ma} is one of the promising practical solutions to BB84 QKD with coherent state light pulses, in which several coherent state light pulses with different intensities are used.
Such optical pulses with different intensities have different photon number statistics.
This simple fact equips Alice and Bob with a countermeasure against Eve.
The original idea of the decoy state QKD is due to Hwang \cite{Hwang}.
So far, several experimental demonstrations of decoy state QKD have been reported \cite{ZQMLQ,Fiber,Free-Space,PZYGMYZYWP,YSS}.
In most cases, the security analysis is based on the GLLP's asymptotic arguments, whereas, in the practical setting, the code length is finite so that the asymptotic argument is no longer valid and the {\it unconditional security} is actually not guaranteed any more.
The security analysis of QKD with the finite code size must incorporate the statistical fluctuations of the observed quantities \cite{Hayashi_PRA}.
Although several authors \cite{Wang,Ma,Fiber,HEHN} have considered the influence of statistical fluctuations on the decoy state QKD with finite code length, 
what all of them have done is limited to the re-adjustment of parameters of the asymptotic GLLP's formula for the secure key generation rate.
%Yet, they are still on the level of the modification of GLLP arguments.
Such an {\em ad hoc} treatment cannot be justified to claim the unconditional security.

In this paper, we propose a substantially improved decoy state QKD in the framework of finite coding length \cite{Hayashi_Tight} and report on its experimental results.
In our experiment, we employ the decoy state method with three decoy
pulses and demonstrate that in the 20 km optical fiber transmission, the
final key was successfully generated at the rate 200 bps keeping
the eavesdropper's mutual information with the final key less than $2^{-9}$.

In our protocol, we use $k+1$ different intensities or mean photon numbers $\mu _{0}=0<\mu _{1}<\ldots <\mu _{k}$ including vacuum ($\mu _{0} $) for the optical pulses.
Two conjugate bases ($+$ and $\times $) are treated separately so that $2k+1$ different pulses are involved in total.
The vacuum state ($i=0$) is sent at the probability $\overline{p}_{0}$ and the $\mu _{i} $ pulse with $\times $ ($+$) basis is sent at the probability $\overline{p}_{i}$ ($\overline{p}_{i+k}$) ($i=1,\ldots ,k$)
The pulse with intensity $\mu _{i_{0}}$ ($\mu _{i_{0}+k}$) (the signal pulse) is used to distill the final secret key and the remainings (decoy pulses) are used just for estimation of Eve's attacks and/or the noise characteristics of quantum channel.

The code length (size of raw key) is denoted by $N$
and the time slot to
generate a final secure key is denoted by $T$.
We also fix the maximum number $\overline{N}$ of the size of final key.
Our protocol is as follows.
Within the time slot $T$, Alice randomly sends Bob a sequence of optical pulses of $k+1$ different intensities with randomly chosen basis.
After that, Bob performs a measurement in one of the two bases and Alice and Bob compare bases and keep the pulses with a common basis by communicating via public channel.
The number of sending pulses, received pulses, and pulses of a common basis are denoted by, respectively, $A_{i}$, $C_{i}$, and $E_{i}$ ($i=0,\ldots, 2k$).
The $E_{i} $ bit string of $i$th pulse contains error bits, which will be detected by checking a portion of the bits (check bits).
To prepare check bits, Alice and Bob firstly perform the random permutation on $E_{i_{0}} $ and $E_{i_{0}+k} $ bit strings by sharing common random numbers via public channel.
Then, for $i=i_{0}$ and $i=i_{0}+k$, the first $N$ bit string is used as the raw key and the remaining $E_{i_{0}}-N$ and $E_{i_{0}+k}-N$ bit string are used as the check bits, while the whole $E_{i}$ bits are used as check bits for $i\neq i_{0},i_{0}+k$.
(If $E_{i_{0}}\leq N$ or $E_{i_{0}+k}\leq N$, then the protocol is aborted.)
The number of detected errors of $i$th pulse is denoted by $H_{i}$ $(i=1,\ldots,2k)$.
From these quantities, Alice and Bob can evaluate the size of the final key guaranteeing the unconditional security.
If the evaluated final key size is not positive, the protocol is aborted again.
The final secret key $N_{final}$ is computed as 
$
N\eta \left( H_{i_{0}+k}/(E_{i_{0}+k}-N)\right) -m_{max}
$ 
for $+$ basis and $\tilde{N}_{final}$ is computed as
$
N\eta \left( H_{i_{0}}/(E_{i_{0}}-N)\right) -\widetilde{m}_{max}
$ 
for $\times$ basis, where $\eta (\cdot )$ denotes the error correcting coding rate and 
$m_{max}$ ($\widetilde{m}_{max}$) represents the size of privacy
amplification.
If $\overline{N} < N_{final}$ ($\overline{N} < \tilde{N}_{final}$),
they replace $N_{final}$ ($\tilde{N}_{final}$) by $\overline{N}$.

The error correction (or reverse reconciliation in our protocol) is performed as follows.
Suppose that Alice and Bob have, respectively, the random number sequences $X$ and $X^{\prime }$ of $n$ bits, which contain some errors.
The task is to distill the common random number sequence of $l$ bits with negligible errors.
Let $\mathbf{G}$ be the generator matrix of $[n,l]$ classical error correcting code.
Bob generates the random number sequence $Z\in \{0,1\}^{l}$ and sends the bit string 
$\mathbf{G}Z\mathbf{+}X^{\prime }$ to Alice.
Then Alice decodes $\mathbf{G}Z\mathbf{+}X^{\prime }-X$ to extract $Z$.
%This process involves the twirling so that the quantum channel from Bob to Alice is essentially a Pauli channel \cite{Hayashi_PRA}.
For the classical error correcting code, Low Density Parity Check (LDPC) code \cite{LDPC} is used.
The advantage of LDPC code is that the decoding with $O(n)$ operations is possible by using Sum-Product decoding method \cite{LDPC}, where $n$ is the coding length.
Furthermore, the coding rate achieves the Shannon limit asymptotically.
In the privacy amplification, we use the universal$_{2}$ hash function.
More specifically, we use $(l-m) \times l$ Toeplitz matrix $M_{p}$ \cite{Toeplitz} to subtract $m$ bit information from the original information of $l$ bits.

Now, let us describe the Eve's possible attacks.
To this end, we firstly define multiphoton states $\rho _{l}$ ($l=2, \ldots, k+1$) as 
$
\rho _{l}=\Omega _{i}^{-1}\sum_{n=i}^{\infty }\frac{\gamma _{l,n}}{n!}%
\left| n\right\rangle \left\langle n\right| 
$ with
\[
\gamma _{l,n}=\sum_{j=1}^{l-1}\frac{(\mu _{l-1}-\mu _{l-2})\cdots (\mu
_{l-1}-\mu _{1})\mu _{l-1}^{2}\mu _{j}^{n-2}}{(\mu _{j}-\mu _{l-1})\cdots
(\mu _{j}-\mu _{j+1})(\mu _{j}-\mu _{j-1})\cdots (\mu _{j}-\mu _{1})},
\]
$\Omega _{l} $ being the normalization constant and $\mu _{1}<\mu _{2}<\ldots <\mu _{k}$.
The phase-randomized coherent state, 
$e^{-\mu _{i}}\sum_{n=0}^{\infty }\frac{\mu _{i}^{n}}{n!}\left|
n\right\rangle \left\langle n\right| $
can be expressed as a convex combination of 
$\left| 0\right\rangle \left\langle 0\right| $, 
$\left| 1\right\rangle \left\langle 1\right| $, and $\rho _{i}$.
Here, we adopt the worst case scenario.
Namely, we assume that Eve can distinguish vacuum state ($j=0$), single photon state ($j=1$), multiphton states $\rho _{2}, \ldots, \rho _{k+1}$ with $\times $ basis ($j=2, \ldots, k+1$) and those with $+ $ basis ($j=k+2, \ldots, 2k+1$).
The number of $j$th state ($j=0, \ldots, 2k+1$) is denoted by $B^{j}$.
According to the values of $B^{j}$, Eve can do the following attacks;
She tricks Bob into detecting the $j$th state with probability $q^{j}$ and causes phase errors with probability $r^{j}$ for the $j$th state ($j=1,2,\ldots,k+1$) and bit errors with probability $\widetilde{r}^{j}$ for the $j$th state ($j=1,k+2,\ldots,2k+1$).
In the following, we focus on the $+$ basis case.
The detection ratio 
$p_{i}=C_{i}/A_{i}$ $(i=0,\ldots ,2k)$
is written in terms of $q^{j}$ as
\begin{equation} \label{eq:detection}
p_{i}=\sum_{j=0}^{2k+1}P_{i}^{j}q^{j}+p_{D},
\end{equation}
where $p_{D}$ is the detector dark count rate and $P_{i}^{j}$ is the generation probability of the $j$th state given that the $i$th pulse is emitted.
The error probability 
$s_{i}=H_{i}/E_{i}$ $(i\neq i_{0},1\leq i\leq k)$ and $s_{i_{0}}=H_{i_{0}}/(E_{i_{0}}-N)$ 
satisfies \cite{Comment}
\begin{equation} \label{eq:error}
s_{i}p_{i}=\sum_{j=1}^{k+1}P_{i}^{j}q^{j}r^{j}+\frac{1}{2}%
(P_{i}^{0}q^{0}+p_{D}).
\end{equation}

In our decoy state method, Alice and Bob try to estimate parameters $q^{j}$ and $r^{j}$ ($\widetilde{r}^{j}$) from the observed quantities $\mathbf{C}$, $\mathbf{E}$, and $\mathbf{H}$ 
to the best of their ability.
The complete determination is, however, beyond their ability so they put the safety standards most stringent.
The computed size of privacy amplification is thus given by
\begin{equation} \label{eq:PA_1}
m_{max}=\max_{0\leq x\leq \sqrt{2}(1-p_{D}),0\leq y\leq 1}m(x,y),
\end{equation}
where
\begin{eqnarray}
&&m(x,y)  \nonumber  \label{eq:PA_2} \\
&=&m_{\infty }-\frac{Nq^{1}(x)\left[ \overline{h}(r^{1}(x,y))-1\right] }{%
C_{i_{0}+k}}\sqrt{A_{i_{0}+k}P_{i_{0}+k}^{1}(1-P_{i_{0}+k}^{1})}  \nonumber
\\
&&\times \left[ -\Phi ^{-1}(2^{-\delta _{1}})\right]   \nonumber \\
&&+\sqrt{v_{x,y,i_{0}}(\mathbf{q}_{ML}(x,y),%
\mathbf{r}_{ML}(x,y))} \left[ -\Phi ^{-1}(2^{-\delta _{2}})\right]
+\delta _{3}%, \\
%&&\delta_1 = \delta + \lceil \log_2 \overline{N} \rceil + 1, \notag \\
%&&\delta_2 = \delta_3 = \delta + \lceil \log_2 \overline{N} \rceil + 2. \notag
\end{eqnarray}
with
$\delta _{1(2,3)}$ being security parameters.
%$\delta$ being a security parameter.
Here,
\[
\Phi (x)=\frac{1}{\sqrt{2\pi }}\int_{-\infty }^{x}e^{-\frac{x^{2}}{2}}dx
\]
and
\[
\overline{h}(x)=\left\{ 
\begin{array}{cc}
1 & \mbox{if $1/2<x\leq 1$,} \\ 
-x\log _{2}x-(1-x)\log _{2}(1-x) & \mbox{if $0\leq x\leq 1/2$}.
\end{array}
\right. 
\]
%In the actual computation, $\overline{h}(x) $ for small $x$ is replaced by 
%$\overline{h}(a)+(x-a)\overline{h}^{\prime }(a)$ 
%to circumvent the singularity of $\overline{h}(x)$ at $x=0$.
%We choose $a=0.01$ here.
%For more detailed discussion on the choice of parameter $a$, see \cite{HHHT}.
In Eq.~(\ref{eq:PA_2}), 
$x=(q^{k+1}+q^{2k+1})/\sqrt{2}$ and
$y=r^{k+1}$; 
$q^{1}(x)$ and $r^{1}(x,y)$ are direct solutions of Eqs.~(\ref{eq:detection}) and (\ref{eq:error}) as a function of $x$ and $y$, 
while $\mathbf{q}_{ML}(x,y)$ and 
$\mathbf{r}_{ML}(x,y)$ are values of maximal likelihood estimation.
Furthermore, $m_{\infty }$ and $v_{x,y,i_{0}}$ are, respectively, the mean and variance of the stochastic variable 
$
F_{i_{0}+k}^{1}\left[ \overline{h}_{a}\left(
G_{i_{0}+k}^{1}/F_{i_{0}+k}^{1}\right) -1\right] +N-F_{i_{0}+k}^{-1},
$ 
where
$
G_{i}^{j}=r^{j}F_{i}^{j}+\Delta ^{^{\prime }}G_{i}^{j}
$, 
$
F_{i_{0}}^{j}=\frac{E_{i_{0}}-N}{E_{i_{0}}}E_{i_{0}}^{j}+\Delta ^{^{\prime
}}F_{i_{0}}^{j}
$ and 
$
F_{i_{0}+k}^{j}=\frac{N}{E_{i_{0}+k}}E_{i_{0}+k}^{j}+\Delta ^{^{\prime
}}F_{i_{0}+k}^{j}
$
with
$
E_{i}^{j}=\frac{1}{2}C_{i}^{j}+\Delta ^{^{\prime }}E_{i}^{j}
$, 
and 
$
C_{i}^{j}=q^{j}B_{i}^{j}+\Delta ^{^{\prime }}C_{i}^{j}
$ 
($
C_{i}^{-1}=p_{D}A_{i}+\Delta ^{^{\prime }}C_{i}^{-1}
$). 
These stochastic variables as well as $H_{i}$ obey the respective hypergeometric distributions, which are fully incorporated in the computation of $v_{x,y,i_{0}}$.

%Choosing $m$ [Eq.~(\ref{eq:PA_1})] as the size of privacy amplification,
%the leakage information is estimated as $\leq 2^{-\delta}$,
%where $\delta_1 = \delta + \lceil \log_2 \overline{N} \rceil + 1$ and
%$\delta_2 = \delta_3 = \delta + \lceil \log_2 \overline{N} \rceil + 2$.
%The detail will be published elsewhere \cite{HHHT}.
Now, let us go back to Eq.~(\ref{eq:PA_1}).
To ensure that the leakage information is less than $2^{-\delta}$,
it is sufficient to choose
$\delta_1 = \delta + \lceil \log_2 \overline{N} \rceil + 1$ and
$\delta_2 = \delta_3 = \delta + \lceil \log_2 \overline{N} \rceil + 2$.
The detail will be published elsewhere \cite{HHHT}.

%%%%%%%%%%%%%%%%%%%%%%%%%revised by AT%%%%%%%%%%%%%%%%%%%%%%%%%%%%
{\em Experiment.---}
The key generation experiment was done with a 20 km-long optical fiber in a common office environment. 
The setup, working on the 62.5 MHz clock, was based on the "plug-and-play" QKD system used in the 14-days-continuous quantum key distribution experiment \cite{TMTT}. 
The wavelength of the light was 1.55 $\mu$m.
The intensity of the optical pulses were randomly chosen with a dual-drive Lithium Niobate Mach-Zehnder intensity modulator \cite{TTMTT06}. 
The modulator provides phase shifts simultaneously to define the four BB84 states, as the output wave is given by 
$
E_{out}=E_{int} \cos[(\phi_1-\phi_2)/2] \exp[i (\phi_1+\phi_2)/2],
$
where $\phi_1$ and $\phi_2$ are the phase shifts in the two arms of Mach-Zehnder interferometer.
We employed  "alternative-shifted-phase-modulation" to use a polarization dependent device \cite{T5N04}.
A PC with Pentium(R)4 (3GHz) CPU and 2 GB memory was connected to the QKD apparatus to perform the error correction and privacy amplification.
The random numbers used in the experiment were generated by a physical random number generation hardware. %with the generation rate 250 kbps.
%%%%%%%%%%%%%%%%%%%%%%%%%%%%%%revise end%%%%%%%%%%%

Before running the protocol, the detector dark count rate $p_{D}$ must be determined in advance, which was measured as $3.00\times 10^{-4}$ per pulse.
We used pulses with three different finite intensities $\mu _{1}$, $\mu _{2}$, and $\mu _{3}$ in addition to the vacuum state ($k=3$) because at least four different intensities including vacuum are needed to achieve the almost optimal secure key generation rate asymptotically \cite{Hayashi_Asympt}.
The final key was distilled from the $\mu _{3}$ pulses sequence.
We set $\mu _{3} = 0.5$.
This is due to the observation that at the limit $\mu _{1},\mu _{2} \rightarrow 0$, 
the choice of $\mu _{c}=0.5$ yields the best final key generation rate asymptotically \cite{Hayashi_Asympt}.
The others are set to be $\mu _{1} = 0.07$ and $\mu _{2} = 0.35$, which are the smallest two values available in our system.
The security parameters were set to be $\delta=9$ so that Eve's mutual information with the final key is guaranteed to be less than $2^{-9}$~\cite{HHHT}.

The size of LDPC code was $1.0 \times 10^{4}$.
The Sum-Product method is an approximate decoding one so that there is a finite probability of decoding without converging.
In our case, the unconverging probability was about $3 \times 10^{-3}$.
If the decoding is unsuccessful, the bit string must be discarded, but after the successful decoding, the bit error rate was reduced to as small as $1.0 \times 10^{-10}$.
The coding length $N$ was set to be 10 times of the LDPC coding length $1.0 \times 10^{4}$ ($N = 1.0 \times 10^{5}$) and the error correction was performed on each $10^{4}$ bit block of $N $.

\begin{figure}
 \begin{center}
  \includegraphics[width=0.49\textwidth]{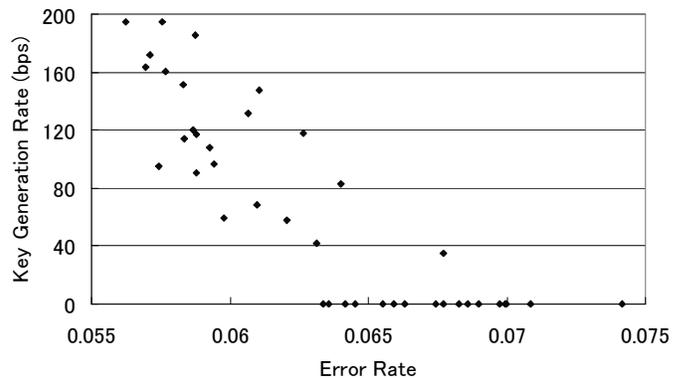}
 \end{center}
 \vspace{-2em}
 \caption{\label{fig:fig1}Sum of key generation rates of $+$ and $\times
 $ bases. Error rates were the average of two bases.
 }
\end{figure}

\begin{table}
\caption{\label{tab:table1}Number of received bits and check bits.}
\begin{ruledtabular}
\begin{tabular}{crrrr}
& \multicolumn{2}{c}{Received Bits} & \multicolumn{2}{c}{Check Bits} \\ \hline
Intensity & $+$ basis & $\times $ basis & $+$ basis & $\times $ basis \\ \hline
Vacuum & \multicolumn{2}{c}{52399} \\
0.07 & 172935 & 173779 &  88406 &  84430 \\
0.35 & 177279 & 178666 &  89967 &  87700 \\
0.50 & 784750 & 786163 & 294847 & 292321 \\
\end{tabular}
\end{ruledtabular}
\end{table}

We performed the above-described QKD experiment for 40 rounds.
The sum of the key generation rate of $+$ and $\times $ bases is shown
in Fig~\ref{fig:fig1}.
We set the time slot $T=41.8$ sec and 
$A_{0}:A_{1}(=A_{4}):A_{2}(=A_{5}):A_{3}(=A_{6})=0.125:0.1875:0.0625:0.1875$.
We also fix $\overline{N}=2^{12}$ which maximizes the average of the key
generation rates.
The number of received pulses $C_{i}$ and the check bits for the coding
length $N = 1.0 \times 10^{5}$ are listed in Table~\ref{tab:table1}.
When the error rate on $+$ ($\times$) basis was $5.2\%$ ($6.1\%$),
the final secret key of $8.2 \times 10^{3}$ bit was generated and
the generation rate was around 200 bps.
In this case the raw key of $N =1.0 \times 10^{5}$ bit was reduced to the size of
$5.6 \times 10^{4}$ bit by error correction and further to the size of
$4.1 \times 10^{3}$ bit by privacy amplification on each basis.
%The key generation rate obtained by our decoy method looks so small.
%However the security of the final key can not be guaranteed
%until so many bits are discard by privacy amplification
%under the strict statistical fluctuations of the observed quantities
%even if the code length is $10\times 10^{5}$.
If the error rates were more than $6.5 \%$,
then few final secret key was left.
%On average, the final secret key of $2.8 \times 10^{3}$ bit was
%generated and the averaged generation rate was around 67 bps.
Although the key generation rates
obtained in our experiments are not so large, the most important point
is that the final keys are guaranteed to be unconditionally secure
in our decoy method while previously reported ones
\cite{ZQMLQ,Fiber,Free-Space,PZYGMYZYWP,YSS} are not.
The small values of the key generation rates are due to the statistical
fluctuations of observed quantities which are never negligible even if
the code length is $1.0\times 10^5$.

{\em Discussion on experimental parameters.---}
Many adjustable parameters are involved in our protocol and some of them are in the trade-off relation.
In the following, we clarify, albeit qualitatively, the dependence of these parameters on our decoy state QKD experiment for further improvements.

Firstly, the sending probability of the signal pulse must be large enough to ensure $E_{3}, E_{6} \geq N$.
We also performed the same experiment but under the condition 
$A_{0}:A_{1}(=A_{4}):A_{2}(=A_{5}):A_{3}(=A_{6})=0.125:0.125:0.125:0.125$ 
and the resulting generation rate of the secret key was around 50 bps,
which is less than that of Fig~\ref{fig:fig1}.
However, the sending probability of pulses with intensities $\mu _{1}$ and $\mu _{2}$ also must not be too small.
Otherwise, the statistical error of estimating Eve's parameters would become large reducing the size of final key.
After all, our choice of sending probabilities
that check bits for the intensity 0.07 are as much as those for
the intensity 0.35 in Table~\ref{tab:table1} is quite favorable,
although it is not optimized.

Secondly, the processing time of each step in our protocol affects directly the key generation rate.
The most time-consuming process is the sharing of common random numbers via quantum channel.
It took around 42 sec in the experiment to obtain the results of Fig~\ref{fig:fig1}.
The second one is the computation of the size of privacy amplification, which requires the nested numerical optimization and took around 10 sec.
The generation of physical random numbers also took around 10 sec, but they did not degrade the total performance since they were performed in a parallel manner.
The random permutation, error correction, and privacy amplification were the least time-consuming processes; they took only a few seconds.

Thirdly, the length of the time slot $T$ is also one of the adjustable parameters.
The numbers of received pulses $C_{i}$, pulses with a common basis, and the check bits are proportional to $T$.
When these numbers are too small the contribution of statistical fluctuation of the right-hand side of Eq.~(\ref{eq:PA_2}) becomes large reducing the size of the final secret key.
Thus, $T$ must be large enough.
However, too large $T$ merely results in low key generation rate since the parameter estimation is not improved any longer for sufficiently large value of $T$.
Thus, we can expect there is some optimal value of $T$.
%although the actual estimation of the optimal $T$ is a hard problem 
%because it depends on many experimental parameters such as 
%intensities of sending pulses, transmission rate of the quantum channel, and the error rates, etc.
Under the same conditions of our experiment, the largest key generation
rate is expected to be achieved when the smallest size of check bits
(for the intensity 0.07 in Fig~\ref{fig:fig1}) is around 85$\%$
of the coding length $N = 1.0 \times 10^{5}$.

Finally, let us briefly discuss the size of random numbers used in our experiment.
The random numbers, which are all physical random numbers, are used for random permutation, error correction, and the generation of Toeplitz matrix.
Among them, the random permutation is most demanding; it requires the random numbers of the size 
$O(N\log N)\simeq 1.7\times 10^{6}$ bits since each bit in $E_{3}$ ($E_{6}$) is randomly exchanged.

The overall performance of decoy state QKD could be improved if we used the unidirectional QKD system \cite{GYS,unidirectional} instead of 'plug-and-play' system here.
By doing so, a larger bit string of common random numbers would be available so that the better performance of error correction and therefore higher secret key distillation rate would be expected.
In our experiment, the bottleneck of the key generation time is the common random number sharing via quantum channel.
This would be resolved if the transmission rate of optical pulses was improved, however, the computational time of the size of privacy amplification would emerge as the second difficulty.
%Delete if necessary
The time of random number sharing via quantum channel is proportional to the coding length $N $, while the time of generating physical random numbers is proportional to $O(N\log N)$.
Therefore, if $N $ is too large, the generation of random numbers may affect seriously the total performance.
%Delete if necessary

In summary, we have proposed and demonstrated the improved decoy state QKD incorporating finite statistics due to the finite code length.
We employed the decoy state method with four different intensities including vacuum state for optical pulses and achieved the final key generation rate of 200 bps in 20 km telecom optical fiber transmission keeping the eavesdropper's mutual information with the final key as small as $2^{-9}$.
We also discussed the dependence of several parameters on our QKD experiment.

{\em Acknowledgments.---}
We would like to thank Hiroshi Imai for support.
The ``plug-and-play" QKD system used in the experiment was originally
developed by NEC based on research carried out under
the National Institute of Communication Technology's (NICT) project
``Research and Development of Quantum Cryptography."

%The JST team developed the software for implementation in the QKD hardware,
%which was developed by NEC based on research carried out under
%the National Institute of Communication Technology's (NICT) project
%``Research and Development of Quantum Cryptography."
%The "plug-and-play" QKD system used in the experiment was originally developed by NEC Corp. and National Institute of Communication Technology, Japan.

%%%%%%%%%%%%%%%%%%%%%%%%%%%%%%%%%%%%%%%%%%%%%%%%%%%%%%%%%%%%%%%%%%%%%%%%%%%%%%%
%%
%%%%%%%%%%%%%%%%%%%%%%%%%%%%%%%%%%%%%%%%%%%%%%%%%%%%%%%%%%%%%%%%%%%%%%%%%%%%%%%

%%%%%%%%%%%%%%%%%%%%%%%%%%%%%%%%%%%%%%%%%%%%%%%%%%%%%%%%%%%%%%%%%%%%%%%%%%%%%%%
%%
%%%%%%%%%%%%%%%%%%%%%%%%%%%%%%%%%%%%%%%%%%%%%%%%%%%%%%%%%%%%%%%%%%%%%%%%%%%%%%%


\begin{thebibliography}{99}

\bibitem{BB84}
C. H. Bennett and G. Brassard, 
in {\it Proceedings of IEEE International Conference on Computers, Systems, and Signal Processing, Bangalore, India, 1984} (IEEE, New York, 1984), p. 175.

\bibitem{QKD_Review}
N. Gisin {\it et al.}, 
%N. Gisin, G. Ribordy, W. Tittel, and H. Zbinden, 
Rev. Mod. Phys. {\bf 74}, 145 (2002).

\bibitem{Security}
D. Mayers, in {\it Advances in Cryptology --- Proceedings of Crypto '96};
Lecture Notes in Computer Science, {\bf 1109}, 343 (1996); 
J. ACM {\bf 48}, 351 (2001); 
H.-K. Lo and H. F. Chau, Science {\bf 283}, 2050 (1999); 
P. W. Shor and J. Preskill, Phys. Rev. Lett. {\bf 85} 411 (2000); 
H. Inamori, N. L\"{u}tkenhaus, and D. Mayers, quant-ph/0107017.

\bibitem{PNS}
B. Huttner {\it et al.}, 
%B. Huttner, N. Imoto, N. Gisin, and T. Mor, 
Phys. Rev. A {\bf 51}, 1863 (1995); 
G. Brassard {\it et al.}, 
%G. Brassard, N. L\"{u}tkenhaus, T. Mor, and B. Sanders, 
Phys. Rev. Lett. {\bf 85}, 1330 (2000); 
N. L\"{u}tkenhaus and M. Jahma, New J. Phys. {\bf 4} 44 (2002).
%Quantum key distribution with realistic states: photon-number statistics in the photon-number splitting attack

\bibitem{GLLP}
D. Gottesman {\it et al.}, 
%D. Gottesman, H.-K. Lo, N. L\"{u}tkenhaus, and J. Preskill, 
Quantum Inf. Comput. {\bf 4}, 325 (2004).

\bibitem{Hwang}
W.-Y. Hwang, Phys. Rev. Lett. {\bf 91} 057901 (2003).

\bibitem{Wang}
X.-B. Wang, Phys. Rev. A {\bf 72} 012322 (2005); Phys. Rev. Lett. {\bf 94} 230503 (2005).
%Decoy-state protocol for quantum cryptography with four different intensities of coherent light
%Beating the Photon-Number-Splitting Attack in Practical Quantum Cryptography

\bibitem{LMC}
H.-K. Lo, X. Ma, and K. Chen, 
Phys. Rev. Lett. {\bf 94} 230504 (2005).
%Decoy State Quantum Key Distribution

\bibitem{Ma}
X. Ma {\it et al.}, 
%X. Ma, B. Qi, Y. Zhao, and H.-K. Lo, 
Phys. Rev. A {\bf 72}, 012326 (2005); 
%X. Ma, C.-Z. F. Fung, F. Dupuis, K. Chen, K. Tamaki, and H.-K. Lo, 
{\em ibid.} {\bf 74} 032330 (2006).
%Practical decoy state for quantum key distribution
%Decoy-state quantum key distribution with two-way classical postprocessing

\bibitem{ZQMLQ}
Y. Zhao {\it et al.}, 
%Y. Zhao, B. Qi, X. Ma, H.-K. Lo, and L. Qian, 
Phys. Rev. Lett. {\bf 96} 070502 (2006).
%Experimental Quantum Key Distribution with Decoy States

\bibitem{Fiber}
D. Rosenberg {\it et al.}, 
%D. Rosenberg, J. W. Harrington, P. R. Rice, P. A. Hiskett, C. G. Peterson, R. J. Hughes, A. E. Lita, S. W. Nam, and J. E. Nordholt, 
Phys. Rev. Lett. {\bf 98} 010503 (2007).
%Long-Distance Decoy-State Quantum Key Distribution in Optical Fiber

\bibitem{Free-Space}
T. Schmitt-Manderbach {\it et al.}, 
%T. Schmitt-Manderbach, H. Weier, M. F\"{u}rst, R. Ursin, F. Tiefenbacher, T. Scheidl, J. Perdigues, Z. Sodnik, C. Kurtsiefer, J. G. Rarity, A. Zeilinger, and H. Weinfurter, 
Phys. Rev. Lett. {\bf 98} 010504 (2007).
%Experimental Demonstration of Free-Space Decoy-State Quantum Key Distribution over 144 km

\bibitem{PZYGMYZYWP}
C.-Z. Peng {\it et al.}, 
%C.-Z. Peng, J. Zhang, D. Yang, W.-B. Gao, H.-X. Ma, H. Yin, H.-P. Zeng, T. Yang, X.-W. Wang, and J.-W. Pan, 
Phys. Rev. Lett. {\bf 98} 010505 (2007).
%Experimental Long-Distance Decoy-State Key Distribution Based on Polarization Encoding

\bibitem{YSS}
Z. L. Yuan, A. W. Sharpe, and A. J. Shields, 
Appl. Phys. Lett. {\bf 90} 011118 (2007).

\bibitem{Hayashi_PRA}
M. Hayashi, Phys. Rev. A {\bf 74}, 022307 (2006).

\bibitem{HEHN}
J. W. Harrington {\it et al.}, 
%J. W. Harrington, J. M. Ettinger, R. J. Hughes, and J. E. Nordholt, 
quant-ph/0503002.
%Enhancing practical security of quantum key distribution with a few decoy states

\bibitem{Hayashi_Tight}
M. Hayashi, Phys. Rev. A, to appear; quant-ph/0702250.

\bibitem{LDPC}
R. G. Gallager, MIT Press, Cambridge, MA (1963).
D. J. C. MacKey, IEEE Trans. Inform. Theory {\bf 45}, 399 (1999).

%\bibitem{Sum-Product}

\bibitem{Toeplitz}
L. Carter and M. Wegman, 
J. Computer and System Sciences {\bf 18}, 143 (1979); 
H. Krawczyk, 
in {\it Advances in Cryptology -- CRYPTO '94, 14th International Cryptology Conference}; 
Lecture Notes in Computer Science {\bf 839}, 129 (1994).

\bibitem{Comment}
More correctly, the errors of each basis other than the detector dark count events may occur even when the quantum channel does not cause any errors.
Here, we neglect such error probability.
This assumption is not harmful to the security arguments.
On the contrary, this means that these errors are under the full control of Eve.

\bibitem{HHHT}
J. Hasegawa {\it et al.}, to be submitted to Phys. Rev. A.

\bibitem{TMTT}
A. Tanaka {\it et al.}, 
%A. Tanaka, W. Maeda, A. Tajima, and S. Takahashi, 
in {\it Proceedings of the 18th Annual Meeting of the IEEE Lasers and Electro-Optics Society, Sidney, Australia, 2005} (IEEE, New York, 2005), p. 557.

\bibitem{TTMTT06}
A. Tanaka {\it et al.}, 
%A. Tanaka, A. Tomita, W. Maeda, S. Takahashi, A. Tajima, 
{\it the 32nd European Conf. Optical Commun., Cannes, France, 2006} We3.P.186.

\bibitem{T5N04}
A. Tanaka {\it et al.}, 
%A. Tanaka, A. Tomita, A. Tajima, T. Takeuchi, S. Takahashi, Y. Nambu, 
{\it the 30th European Conf. Optical Commun., Stockholm, Sweden, 2004}, Tu4.5.3.

\bibitem{Hayashi_Asympt}
M. Hayashi, 
quant-ph/0702251.

\bibitem{GYS}
C. Gobby, Z. L. Yuan, and A. J. Shields, 
Appl. Phys. Lett. {\bf 84} 3762 (2004).

\bibitem{unidirectional}
T. Kimura {\it et al.}, 
%T. Kimura, Y. Nambu, T. Hatanaka, A. Tomita, H. Kosaka, and K. Nakamura, 
Jpn. J. Appl. Phys. {\bf 43} L1217 (2004).

\end{thebibliography}
\end{document}